\documentclass[tightenlines,showpacs,prl,
floatfix,preprintnumbers,amsmath,amssymb,amsbsy,twocolumn]{revtex4}
\usepackage{graphicx}
\usepackage{bm}
\begin{document}
\vspace{1cm}

\title{Anisotropies in the Gravitational-Wave Stochastic Background}

\author{S. \"Olmez$^a$, V. Mandic$^a$ and X.~Siemens$^b$}
\affiliation{$^a$School of Physics and Astronomy, University of Minnesota, Minneapolis, MN 55455\\
$^b$Center for Gravitation and Cosmology, University of Wisconsin - Milwaukee, WI 53201}

\date{\today}

\begin{abstract}
We  consider anisotropies in the stochastic background of
gravitational-waves (SBGW) arising from random fluctuations in the
number of  gravitational-wave sources. We first develop the general
formalism which can be applied to different cosmological or
astrophysical scenarios. We then apply this formalism to calculate
the anisotropies of SBGW associated with the fluctuations in the
number of cosmic string
 loops, considering both cosmic string cusps and kinks. We calculate the anisotropies as a function
 of angle and frequency.
\end{abstract}


\bibliographystyle{plain}
\maketitle

\pagestyle{plain} A stochastic background of gravitational-wave
(SBGW) radiation is produced by a large number of weak, independent
and unresolved gravitational wave sources. The sources of the SBGW
can be isotropic or anisotropic. For the case of sources of
cosmological origin \cite{starobinskii, barkana, apreda} the
distribution of the gravitational wave sources  is expected to be
isotropic, while astrophysical sources such as rotating neutron
stars \cite{rotNS} or magnetars \cite{magnetars} may have an
anisotropic distribution. Even in the case of an a priori isotropic
source distribution, random fluctuations in the number of sources
will (in general) give rise to anisotropies. Such anisotropies are
analogous to the anisotropies observed in the cosmic microwave
background (CMB) radiation and would carry additional information
about the gravitational-wave sources that generated them.\\
Cosmic strings are expected to contribute to the SBGW. Cosmic
strings are  predicted by a large class of unified theories
\cite{kibble76,alexbook,Jeannerot:2003qv}
 as remnants of spontaneously broken symmetries at phase transitions in the early Universe,
as well as in string-theory-inspired cosmological scenarios~\cite{cstrings}.
Once formed, a network of cosmic strings evolves toward
an attractor solution called  the scaling regime, in which
the statistical properties of the network, such as the
average distance between strings and the size of loops at formation,
scale with the cosmic time.   The gravitational interaction of strings is characterized by the dimensionless parameter
$G \mu$, where $G$ is Newton's constant and $\mu$ is the tension. The current CMB bound on
the tension is $G \mu<6.1\times 10^{-7}$ \cite{CMB1,CMB2}. Cosmic string cusps, regions of string that acquire
enormous Lorentz boosts,
are expected to generate large transient gravitational-wave signals~\cite{DV1,DV2,SCMMCR}.
Such individual bursts could be observable by current and planned gravitational wave detectors~\cite{Abbott:2009rr}
  for values of $G \mu$ as
low as $10^{-13}$, which may provide a probe of a certain class of
string theories \cite{cstrings}. A SBGW produced by the incoherent
superposition of cusp  bursts from a network of cosmic strings and
superstrings was considered in \cite{DV3,Siemens Mandic Creighton},
and it was later shown that  kinks contribute
 to the SBGW at the same order as cusps \cite{Olmez Mandic Siemens}. These sources of SBGW are
 also observable by current and planned detectors, for a wide range of the parameter space, see \cite{Siemens Mandic Creighton, Olmez Mandic Siemens}
 and references therein.\\
In this paper, we develop a general formalism to treat SBGW
anisotropies. In particular, we consider two-point correlations in
SBGW between two different directions in the sky, which arise from
random fluctuations in the number of gravitational-wave sources.
While this formalism is applicable to a variety of cosmological and
astrophysical SBGW models (see \cite{extragalactic} and the
references therein), we illustrate it for the specific case of
cosmic (super)string cusps
and kinks. \\
{\it Anisotropies in the SBGW:} We start from formalism
in references \cite{DV2,SCMMCR} and extend it to treat angular
dependence. The energy density of SBGW at frequency $f$
corresponding to sources in the direction $\hat\Omega$  is given by
\begin{equation}\label{background}
    \Omega_{\rm gw}(f,\hat\Omega)\equiv \frac{f} {\rho_c} \frac{d\rho_{\rm gw}(\hat\Omega)}{df},
     \end{equation}
 where $d\rho_{\rm gw}$ is the energy density of
gravitational waves in the frequency range $f$ to $f+df$ and
$\rho_c$ is the critical energy density of the Universe. Let us
assume that sources are characterized by a set of parameters
$\boldsymbol{\zeta}$ - in the case of cosmic strings, redshift $z$
is one such parameter. Therefore $\Omega_{\rm gw}$ is an integral
over the parameter space $\boldsymbol{\zeta}$, which we propose to
discretize as follows:
\begin{eqnarray}\label{background2}
    \Omega_{\rm gw}(f,\hat\Omega) & = & \int d\boldsymbol{\zeta} n(\boldsymbol{\zeta},\hat\Omega)w(f,\boldsymbol{\zeta},\hat\Omega) \nonumber \\
    & \simeq &  \sum_{i}  \Delta(\boldsymbol{\zeta}_i)\,
  n(\boldsymbol{\zeta}_i,\hat\Omega)w(f,\boldsymbol{\zeta}_i,\hat\Omega) \nonumber \\
  & \equiv & \sum_{i}
  N(\boldsymbol{\zeta}_i,\hat\Omega)w(f,\boldsymbol{\zeta}_i,\hat\Omega).
       \end{eqnarray}
Here we assume that the parameter space can be divided into disjoint volumes $\Delta(\boldsymbol{\zeta}_i)$,
centered at $\boldsymbol{\zeta}_i$, whose size is large compared to the correlation length of the number of sources.
 In other words, a statistical fluctuation in the number of sources in one volume would have no implications on the number
 of sources in any other volume.
 We further define $n(\boldsymbol{\zeta}_i,\hat\Omega)$ as the number density of sources
 (i.e number per parameter space volume) in the direction $\hat\Omega$ with the parameter set $\boldsymbol{\zeta}_i$,
  and $w(f,\boldsymbol{\zeta}_i,\hat\Omega)$ as the contribution to $\Omega_{\rm gw}$ of one source at frequency $f$,
   in the direction $\hat\Omega$, and with the parameter set $\boldsymbol{\zeta}_i$.
    We also define $N(\boldsymbol{\zeta}_i,\hat\Omega)\equiv\,
  n(\boldsymbol{\zeta}_i,\hat\Omega) \Delta(\boldsymbol{\zeta}_i)$ as the total number of sources with
  the parameters in the range from $\boldsymbol{\zeta}_i$ to
  $\boldsymbol{\zeta}_i+\Delta(\boldsymbol{\zeta}_i)$ and in the direction $\hat\Omega$.
The contribution of one source is given by
\begin{equation}\label{omega def}
     w(f,\boldsymbol{\zeta}_i,\hat\Omega)\equiv\frac{4\pi^2 f^3}{3 H_0^2}
     h^2(f,\boldsymbol{\zeta}_i,\hat\Omega){\cal
     R}(f,\boldsymbol{\zeta}_i,\hat\Omega),
\end{equation}
 where $h(f,\boldsymbol{\zeta}_i,\hat\Omega)$ is the strain of the gravitational wave with frequency $f$ originating
 from a source with parameters $\boldsymbol{\zeta}_i$ and at
  the line of sight $\hat\Omega$. ${\cal R}(f,\boldsymbol{\zeta}_i,\hat\Omega)$ represents the observable part of the
   gravitational radiation from the source, i.e.   it incorporates the  propagation of the wave in the
   expanding universe as well as possible beaming effects, see \cite{DV1,DV2}.

The  angular dependence of  $N$ can originate from anisotropic
source distribution. Moreover, even in the case of an a priori
isotropic SBGW, random fluctuations in the number of sources will
(in general) give rise to anisotropies. Note that
$N(\boldsymbol{\zeta}_i,\hat\Omega)$ are dimensionless numbers,
which are by construction uncorrelated for different values of the
index $i$.
  Assuming Poisson distribution,
  the statistical fluctuations of $N(\boldsymbol{\zeta}_i,\hat\Omega)$ are of order
  $\sqrt{N(\boldsymbol{\zeta}_i,\hat\Omega)}$.
   The corresponding fluctuation in $ \Omega_{gw}$ is
\begin{equation}\label{delta background2}
    \delta\Omega_{gw}(f,\hat\Omega)=  \sum_i
   \delta N(\boldsymbol{\zeta}_i,\hat\Omega)w(f,\boldsymbol{\zeta}_i,\hat\Omega),
    \end{equation}
The two-point correlation of $\delta\Omega_{gw}(f,\hat\Omega)$ at
two different directions reads
\begin{eqnarray}\label{delta background2 corr}
    {\mathcal C}&\equiv&\left\langle\delta\Omega_{gw}(f,\hat\Omega)\delta\Omega_{gw}(f,\hat\Omega')\right\rangle\\
    &=&\sum_{i,\,j} w(f,\boldsymbol{\zeta}_i,\hat\Omega)w(f,\boldsymbol{\zeta}_j,\hat\Omega')
\left\langle\delta N(\boldsymbol{\zeta}_i,\hat\Omega)\delta
N(\boldsymbol{\zeta}_j,\hat\Omega')\right\rangle.\nonumber
    \end{eqnarray}
Since the fluctuations in the number of gravitational-wave sources
are Poissonian, we propose the following bilinear expectation:
    \begin{equation}\label{delta N corr}
\left\langle\delta N(\boldsymbol{\zeta}_i,\hat\Omega)\delta
N(\boldsymbol{\zeta}_j,\hat\Omega')\right\rangle \sim
N(\boldsymbol{\zeta}_i,\hat\Omega){\mathcal
F}(\gamma,\boldsymbol{\zeta}_i)\,\delta_{ij},
    \end{equation}
where  $\gamma$ is the angle between  $\hat\Omega$ and
$\hat\Omega'$.   ${\mathcal F}$ is a function that incorporates the
correlation properties of the gravitational wave sources. Although
the precise form of ${\mathcal F}$ will depend on the problem at
hand, we can discuss several properties of this function. Firstly,
we expect to see the maximum correlation if the two sources are
close to each other in the physical space as well as the parameter
space. Therefore ${\mathcal F}$ must assume its maximum value at
$\gamma=0$, and it should decrease for larger values of $\gamma.$
Since ${\mathcal F}$ constrains $\gamma$, the angle between
$\hat\Omega$ and $\hat\Omega'$, to small values, we keep only
$\hat\Omega$ at the right hand side of Eq. (\ref{delta N corr}).
This is a good approximation as long as $N$ changes slowly with
$\hat{\Omega}$. Below we will consider an explicit example and
discuss form of ${\mathcal F}$ in more detail. Inserting this
 into Eq. (\ref{delta background2 corr}) gives
\begin{eqnarray}\label{delta background2 corr2}
    {\mathcal C}& =& \sum_{i} \Delta(\boldsymbol{\zeta}_i)n(\boldsymbol{\zeta}_i,\hat\Omega)w^2(f,\boldsymbol{\zeta}_i,\hat\Omega)
    {\mathcal F}(\gamma,\boldsymbol{\zeta}_i) \nonumber \\
& \rightarrow & \int
d\boldsymbol{\zeta}n(\boldsymbol{\zeta},\hat\Omega)w^2(f,\boldsymbol{\zeta},\hat\Omega)
    {\mathcal F}(\gamma,\boldsymbol{\zeta}),
    \end{eqnarray}
    where we take the integral limit of the sum.
This is a general expression applicable to both cosmological and
astrophysical problems in which the correlation properties of the
sources are specified by the function  ${\mathcal F}$.
\begin{figure*}[hbtp]
\includegraphics[width=1.8in]{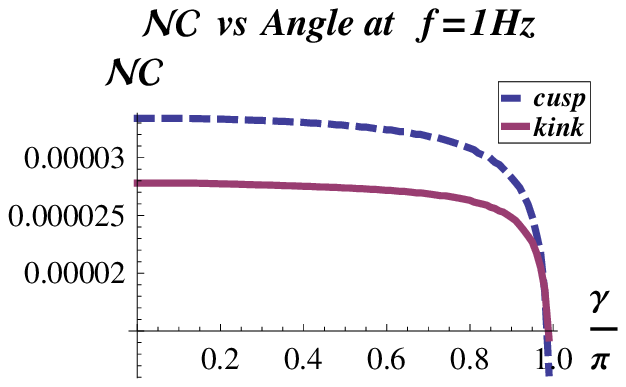}
\includegraphics[width=1.8in]{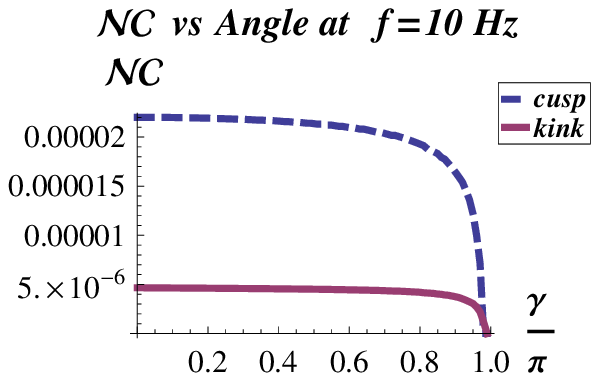}
\includegraphics[width=1.8in]{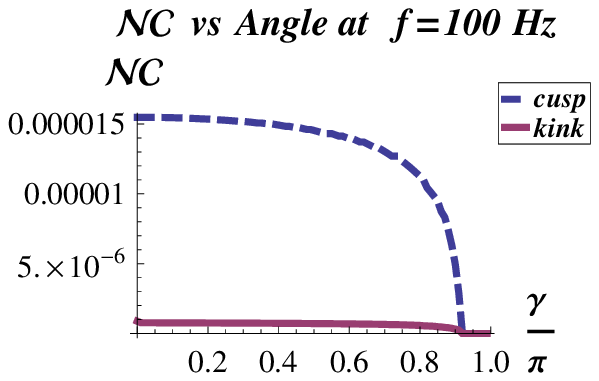}
\includegraphics[width=1.8in]{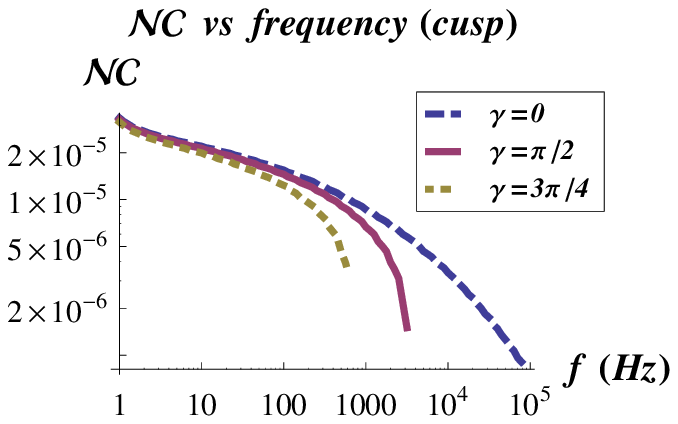}
\includegraphics[width=1.8in]{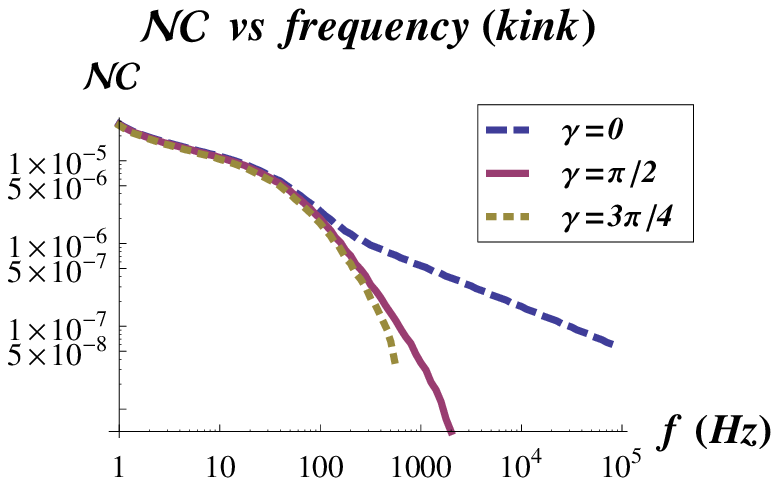}
\caption{Top: Normalized Correlation ${\mathcal N }{\mathcal C }$ as
functions of $\gamma/\pi$ for cusps and kinks for $f=1$Hz, $f=10$Hz
and $f=100$Hz, for  $G\mu=1.0 \times10^{-8}$, $p=1$ and $\epsilon=
1.0 \times   10^{-11}$.\\
 Bottom: ${\mathcal N }{\mathcal C }$  as  functions of frequency for cusps and kinks,
  for various values of $\gamma$ and for the same model parameters as above.} \label{results}
\end{figure*}
\begin{figure*}[hbtp]
\includegraphics[width=1.8in]{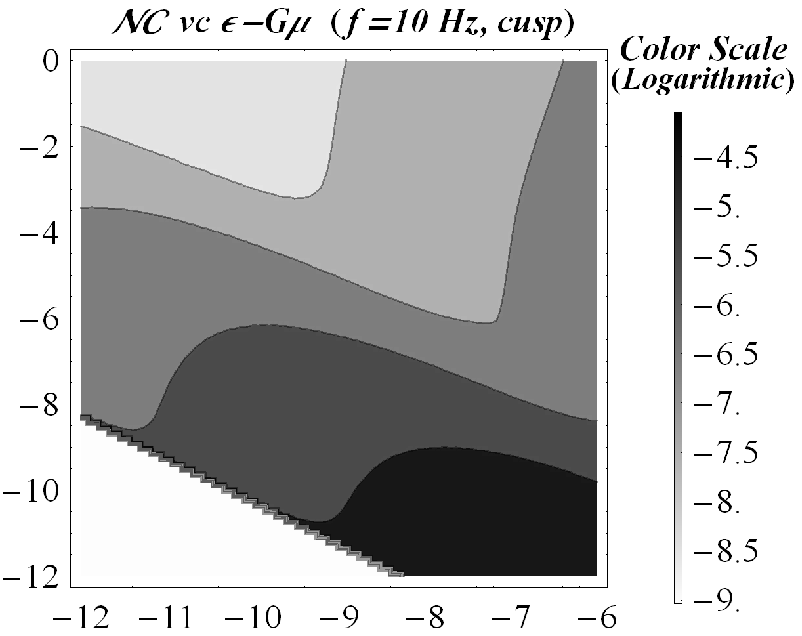}
\includegraphics[width=1.8in]{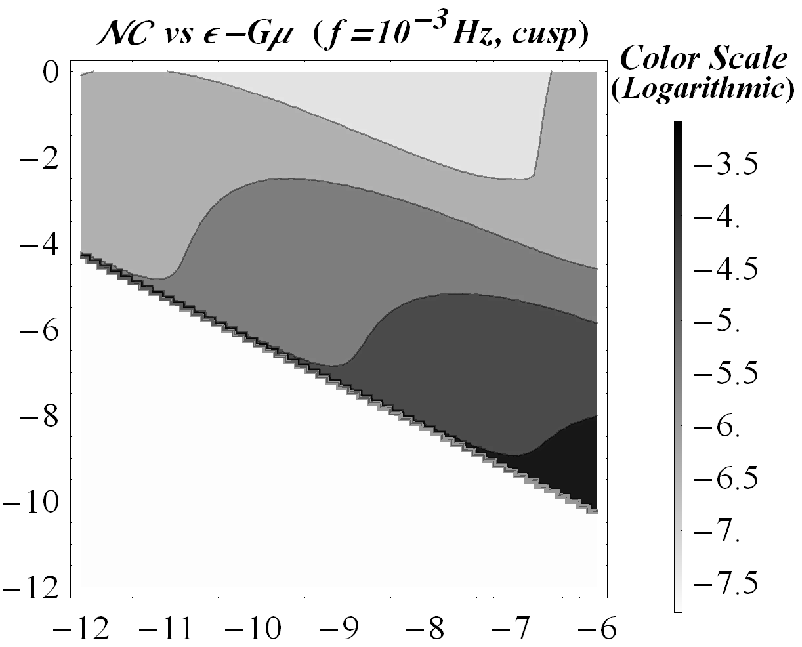}
\includegraphics[width=1.8in]{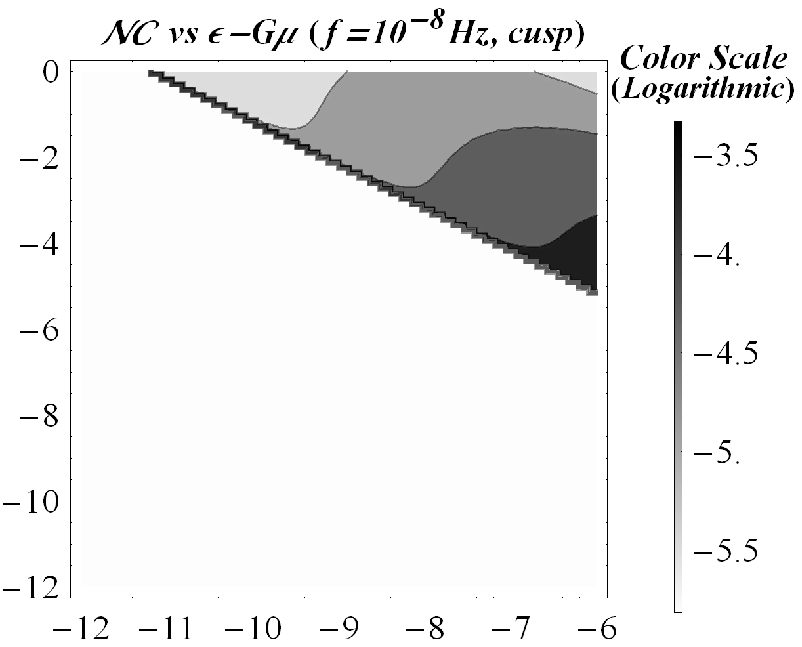}
\includegraphics[width=1.8in]{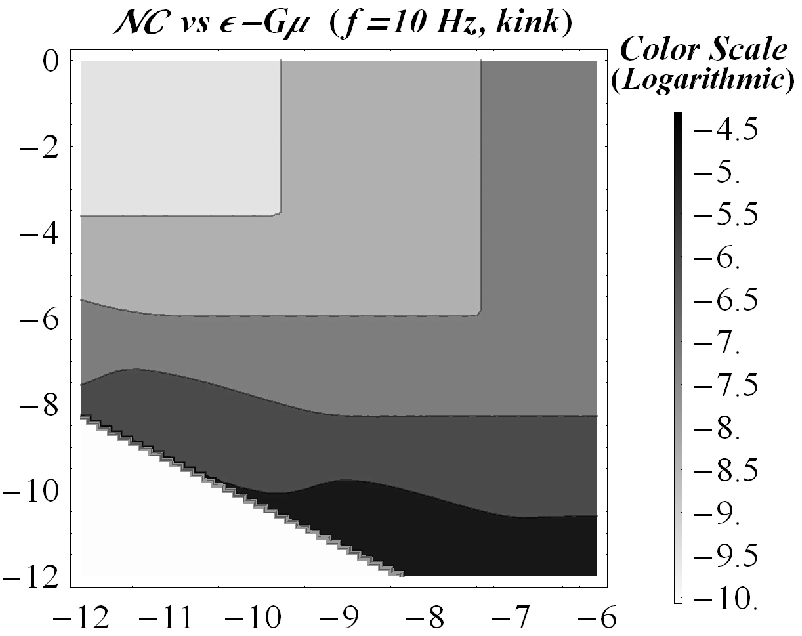}
\includegraphics[width=1.8in]{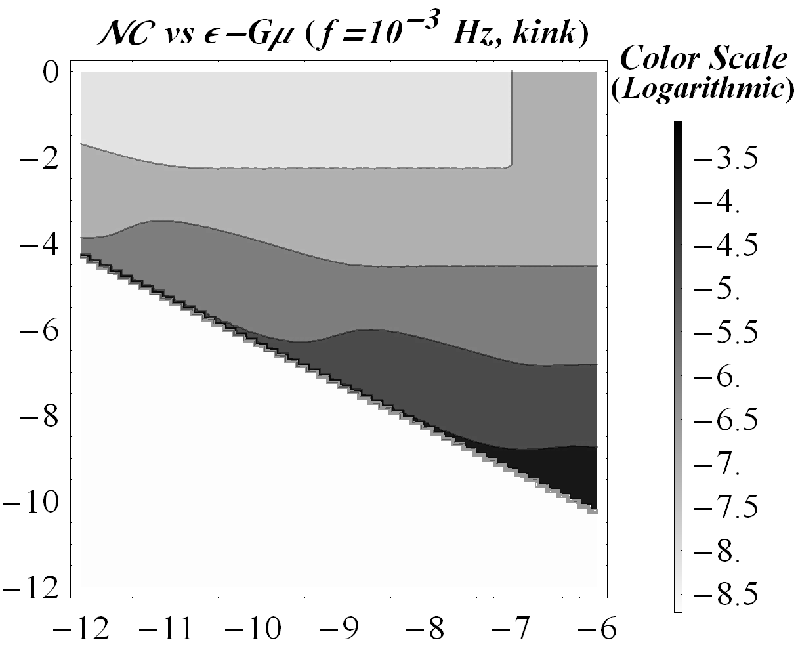}
\includegraphics[width=1.8in]{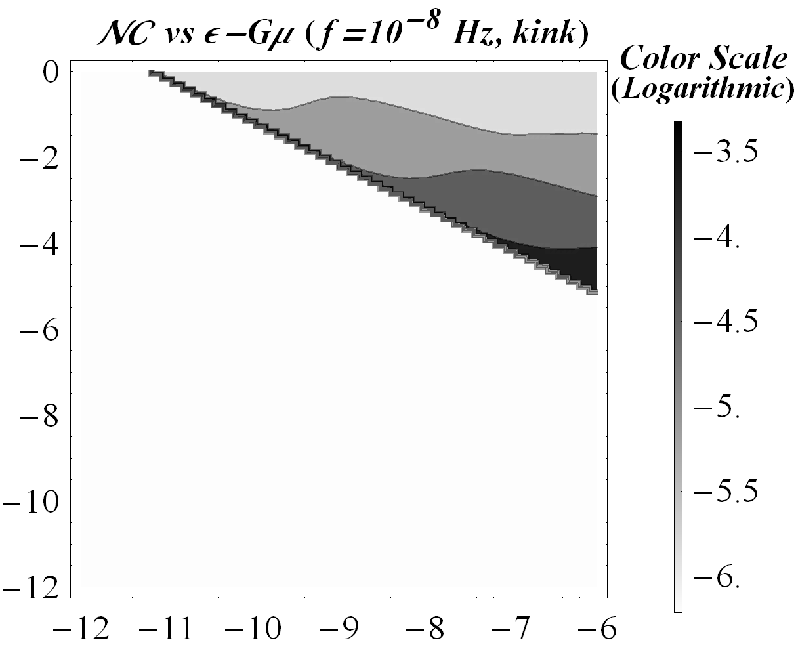}
\caption{${\mathcal N }{\mathcal C }$ for cusps and kinks in
$Log(\epsilon)$ (vertical axis) $Log(G\mu)$(horizontal axis)
parameter space at
 frequencies applicable to ground-based detectors ($10$ Hz) \cite{Abbott:2009rr}, satellite-based
 detectors ($1$ mHz) \cite{Cornish}, and pulsar-based observations ($10^{-8}$ Hz) \cite{Jenet}. The (base $10$ logarithm of)
numerical values of ${\mathcal N }{\mathcal C }$ are denoted in the
color bar for each plot. } \label{results}
\end{figure*}\\

{\it Cosmic Strings Case:} We now apply this formalism to the case
of cosmic strings, including gravitational-wave bursts from cusps
and kinks, in which the distribution of sources is specified by the
redshift $z$. It is convenient to parameterize physical quantities
in terms of redshift. To this end, we define the following
dimensionless cosmological functions:
\begin{eqnarray}\label{phi functions}
\varphi_r(z)&=&\int_0^z\,\frac{dz'}{{\mathcal H}(z')},\nonumber\\
\varphi_t(z)&=&\int_z^\infty\frac{dz'}{(1+z'){\mathcal
H}(z')},\nonumber\\
    \varphi_V(z)&=&\frac{\varphi^2_r(z)}{(1+z)^3{\mathcal
H}(z)},
\end{eqnarray}
where ${\mathcal
H}(z)=\sqrt{\Omega_m(1+z)^3+\Omega_r(1+z)^4+\Omega_\Lambda}$ is the
Hubble function with $\Omega_M=0.25$, $\Omega_R=4.6\times10^{-5}$
and $\Omega_\Lambda=1-\Omega_M-\Omega_R$. We now explicitly
construct the integral in Eq. (\ref{delta background2 corr2}).
Firstly, the parameter space volume $d\boldsymbol{\zeta}$ in this
case is simply the co-moving volume. It can be written as
$H_0^{-3}\varphi_V(z)dz$, where $H_0$ is the present value of the
Hubble constant. This converts the co-moving differential volume
$r^2 dr$ to the corresponding differential volume as a function of
the redshift. The next quantity in Eq. (\ref{delta background2
corr2}) is the number density of the loops. If the loop size is
determined by gravitation back-reaction~\cite{DV1,DV2}, the loop
number density is given by
\begin{equation}\label{loop density}
    n(z) \approx \frac{c(z)}{p \,\Gamma G \mu t^{3}(z)},
\end{equation}
 where $p$ is the reconnection probability, $\Gamma=50$ is
a dimensionless parameter proportional to the power emitted in
gravitational waves by cosmic string loops, and
 $t(z)$ is the cosmic time which can be written
as $t(z)=\varphi_t(z)/H_0$. The function $c(z)\equiv1+\frac{9
z}{z+z_{eq}}$  $(z_{eq}\simeq 5440)$ accounts for the fact that the
loop density in radiation domination is about $10$ times that of the
matter domination. In order to define the ${\mathcal F}$-function in
Eq. (\ref{delta background2 corr2}), we assume that the number
density of cosmic string cusps and kinks at a given $z$ is
correlated over the length scale $R(z)$ given by the Hubble size,
  $R(z)\approx\, t(z)$ \footnote{The factor $(\frac{3}{4\pi})^{\frac{1}{3}}$ is taken as unity.}.
 The angular size spanned by this length scale at the distance
 $r(z)$ can be calculated  using the standard {\it angular diameter-redshift
 relation} as
   \begin{equation}\label{angle}
    \gamma_{z}=2\arctan\left[\frac{(1+z) R(z)}{ r(z)}\right],
 \end{equation}
 where $r(z)$ can be written as $r(z)=\varphi_r(z)/H_0$.
 Therefore, for the given redshift $z$, two directions on the sky are correlated if their angular separation,
 $\gamma$, is less than $\gamma_{z}$. 
 This condition can be imposed by the ${\mathcal F}$ function,
\begin{equation}\label{F theta1}
{\cal F }(\gamma,z)\equiv
\Theta\left[1-\frac{\gamma}{\gamma_{z}}\right],
\end{equation}
 which vanishes if $\gamma$ is larger than $\gamma_z$, the angle subtended by the length scale
 $R(z)$.  We emphasize that the correlations considered here
are large scale, and arise from the fluctuations in the number of
cosmic string loops in an evolving cosmic string network. This is
different from the correlations associated with the correlation
length of a single cosmic string loop, which are important in
determining the cosmic string signatures in the CMB \cite{Yamauchi
et al}.
 For cusps and kinks on cosmic string loops with sizes given by the gravitational back-reaction scale we have \cite{Olmez
Mandic Siemens}
\begin{eqnarray}\label{cusp kink omega}
w_c(f,z)&=& \frac{2 \pi^2  (G\mu)^2}{3 \,(1+z)^{7/3} \varphi_r^2 \,\varphi_t^{1/3} }\frac{\Theta\left[1-(\frac{\alpha f}{H_0}(1+z)\varphi_t)^{-1}\right]}{(\frac{\alpha f}{H_0})^\frac{1}{3}}\nonumber\\
w_k(f,z)&=& \frac{4 \pi^2  (G\mu)^2}{3  \,(1+z)^{8/3} \varphi_r^2
\,\varphi_t^{2/3} }\frac{\Theta\left[1-(\frac{\alpha
f}{H_0}(1+z)\varphi_t)^{-1}\right]}{(\frac{\alpha
f}{H_0})^\frac{2}{3}}, \nonumber \\
&&
 \end{eqnarray}
 where  $\alpha\equiv\epsilon \Gamma G\mu$
 is the parameter that sets the length of the loops.
Since the function $w$ has no angle dependence for kinks and cups,
Eq. (\ref{delta background2 corr2}) simplifies to
\begin{eqnarray}\label{delta background2 corr3}
    {\mathcal C}& = & {\mathcal C}(f,\gamma)=   \int dz\,H_0^{-3}\varphi_V(z) n(z)w^2(f,z)
    {\mathcal F}(\gamma,z) \nonumber \\
    &=& \int dz\varphi_V(z)\frac{c(z)(p \,\Gamma G \mu)^{-1}}{\varphi^3_t(z)}w^2(f,z)\Theta\left[1-\frac{\gamma}{\gamma_{z}}\right]
    \end{eqnarray}
    which is a function of the opening angle, $\gamma$,  and the frequency
    only. It is important to note that  large rare events which occur
at rates smaller than the relevant time-scale of the experiment are
excluded \cite{DV1} from ${\mathcal C}$ in numerical calculation .
This exclusion removes the a priori divergence of the integrand at
$z=0$. The integrand of Eq. (\ref{delta background2 corr3}) quickly
vanishes with increasing redshift, implying that the dominant
contribution comes from low redshifts.
 The    small values of redshift correspond to closer sources, which have
    larger angular size in the sky. Therefore the angular dependence
    of correlations will be rather flat for small angles, and it
    will rapidly vanish for large angles, for which small values of
    redshift are excluded from the integral by ${\mathcal
    F}(\gamma,z)$.
  In order to understand the relative strength
 of the fluctuations at a given frequency $f$
compared to $\Omega_{\rm gw}(f)$ (integrated over all sky) we define
the following quantity: ${\mathcal N }{\mathcal C
}(f,\gamma)\equiv\frac{\sqrt{{\mathcal C}(f,\gamma)}}{\Omega_{\rm
gw}(f)}$, which we refer to as the {\it normalized correlation}. We
numerically evaluate the integrals in Eq. (\ref{delta background2
corr3}) for kinks and cusps and calculate the normalized
correlations, as depicted in Fig. 1. We also do a parameter scan in
$\epsilon-G\mu$ space. Fig. 2 shows the density plot for the
strength of the background $\Omega_{\rm gw}$ and ${\mathcal N
}{\mathcal C }$ at various values of $f$ for
cusps and kinks.\\
{\it Conclusions:} In this paper we have developed the formalism for
calculating the spatial anisotropies in the stochastic background of
gravitational waves associated with the random fluctuations in the
number of sources. The formalism is applicable to a variety of
cosmological and astrophysical models. We applied it to the case of
SBGW due to cosmic (super)string cusps and kinks, and observed that
the relative strength of the anisotropies, $\sqrt{\mathcal C
}/\Omega_{\rm gw}$ , can be estimated by $1/\sqrt N=\sqrt{\Gamma
G\mu}$, which can be as high as $10^{-3}$. While observation of these spatial anisotropies
is unlikely for the second-generation detectors that are currently being built(Advanced LIGO and Advanced Virgo),
 the planned third-generation detectors (such as Einstein Telescope) should be sufficiently sensitive
 to measure them over a large part of the parameter space. We emphasize that the general formalism
 developed here can be used to distinguish between different SBGW models - that is, between models
 that predict similar frequency spectra and different spatial anisotropies. This technique will be
 crucial for the identification of the source of SBGW which is expected to be observed by the future
 generations of the gravitational-wave detectors.\\
 {\it Acknowledgments:} We would like to thank Marco Peloso and Alexander Vilenkin
for useful discussions.  The work of X. S. was supported in part by
NSF grants 0758155 and 0955929 and the work of  V. M. was supported
in part by NSF grant PHY0758036.


\begin{thebibliography} {99}
\bibitem{starobinskii} A.A. Starobinskii, JETP Lett. {\bf 30}, 682 (1979).
\bibitem{barkana} R. Bar-Kana, Phys. Rev. {\bf D 50}, 1157 (1994).
\bibitem{apreda} R. Apreda {\it et al.}, Nucl. Phys. {\bf B 631}, 342 (2002).
\bibitem{rotNS} T. Regimbau and J.A. de Freitas Pacheco, Astron. and Astrophys. {\bf 376}, 381 (2001).
\bibitem{magnetars} T. Regimbau and J.A. de Freitas Pacheco, Astron. and Astrophys. {\bf 447}, 1 (2006).
\bibitem{kibble76} T.W.B. Kibble, J. Phys. {\bf A9}  1387 (1976).
\bibitem{alexbook}
A.~Vilenkin and E.~Shellard, Cosmic strings and other Topological Defects
  (Cambridge University Press, 2000).
\bibitem{Jeannerot:2003qv}
R.~Jeannerot, J.~Rocher, and M.~Sakellariadou, Phys. Rev. \textbf{D
68} 103514 (2003).
\bibitem{cstrings} N. Jones {\it et al.}, JHEP {\bf 0207}
051 (2002); S. Sarangi, S.H.Henry Tye, Phys.Lett. {\bf B536} 185
(2002); G. Dvali, A. Vilenkin, JCAP {\bf0403}  010 (2004); N. Jones
et al., Phys.Lett.  {\bf B563}  6 (2003); E.J. Copeland et al., JHEP
{\bf 0406}  013 (2004), M. Sakellariadou, JCAP {\bf 0504}  003
(2005), J. Polchinski, [arXiv: hep-th/0410082]; J. Polchinski,
[arXiv: hep-th/0412244].
\bibitem{CMB1}L.~Pogosian, M.~C.~Wyman and I.~Wasserman, [arXiv: astro-ph/0403268].
\bibitem{CMB2} E.~Jeong and G.~F.~Smoot, [arXiv: astro-ph/0406432].
\bibitem{DV1}
T.~Damour and A.~Vilenkin,
Phys.\ Rev.\ Lett.\  {\bf 85}, 3761 (2000), [arXiv: gr-qc/0004075].
\bibitem{DV2}
T.~Damour and A.~Vilenkin,
Phys.\ Rev.\  {\bf D64}, 064008 (2001), [arXiv: gr-qc/0104026].
\bibitem{SCMMCR} X. Siemens {\it et al.}, Phys. Rev {\bf D73}  105001 (2006).
\bibitem{Abbott:2009rr}
  B.~P.~Abbott {\it et al.}  [LIGO Scientific Collaboration],
  Phys.\ Rev.\   {\bf D80}, 062002 (2009), [arXiv: astro-ph/0904.4718 ].
\bibitem{DV3}
T.~Damour and A.~Vilenkin,  Phys. Rev. {\bf D71}, 063510 (2005),
[arXiv: hep-th/0410222].
\bibitem{Siemens Mandic Creighton} X. Siemens, V. Mandic and J. Creighton, Phys.\ Rev.\ Lett. {\bf  98}, 111101 (2007).
\bibitem{Olmez Mandic Siemens} S. Olmez, V. Mandic and X. Siemens,  Phys. Rev. {\bf D81} 104028 (2010).
\bibitem{extragalactic} R. Schneider, S. Marassi and V. Ferrari, [arXiv:
astro-ph/1005.0977], Michele Maggiore Phys. Rept. {\bf 331} (2000)
283-367.
\bibitem{Yamauchi et al} D. Yamauchi {\it et al.}, Phys. Rev. {\bf D82}, 063518
(2010).
\bibitem{Cornish} N. J. Cornish, Phys. Rev. {\bf D65} 022004 (2001).
\bibitem{Jenet}F. A. Jenet {\it et al.} , ApJL 625, L123–L126 (2005).
\end{thebibliography}
\end{document}